\documentclass[prb,aps,showpacs,twocolumn,longbibliography,superscriptaddress, ]{revtex4-1}

\usepackage{color} 
\usepackage{bm}
\usepackage{dcolumn}
\usepackage{times}
\usepackage{amssymb} 
\usepackage{amsmath} 
\usepackage{braket}
\usepackage{ragged2e}
\usepackage{graphicx}
\usepackage{epstopdf}
\usepackage[english]{babel}
\usepackage{mathrsfs}
\usepackage{textcomp}
\usepackage{amsthm}
\usepackage{amsfonts}
\usepackage[dvipsnames]{xcolor}
\usepackage{hyperref}
\usepackage{soul}
\usepackage{braket} 
\usepackage{hyperref}
\usepackage{cleveref}
\newcommand{\be}{\begin{equation}}
\newcommand{\ee}{\end{equation}}
\newcommand{\bea}{\begin{eqnarray}}
\newcommand{\eea}{\end{eqnarray}}
\newcommand{\ba}{\begin{array}}
\newcommand{\ea}{\end{array}}

\begin{document}

\title{Topological Phase Detection through High Harmonic Spectroscopy in Extended  Su-Schrieffer-Heeger Chains}
\author{Mohit Lal Bera}
\affiliation{ICFO - Institut de Ci\`encies Fot\`oniques, The Barcelona Institute of Science and Technology, 08860 Castelldefels (Barcelona), Spain.}

\author{Jessica O. de Almeida}
\affiliation{ICFO - Institut de Ci\`encies Fot\`oniques, The Barcelona Institute of Science and Technology, 08860 Castelldefels (Barcelona), Spain.}

\author{Marlena Dziurawiec}
\affiliation{Institute of Theoretical Physics, Wroc{\l}aw University of Science and Technology, 50-370 Wroc{\l}aw, Poland}

\author{Marcin Płodzień}
\affiliation{ICFO - Institut de Ci\`encies Fot\`oniques, The Barcelona Institute of Science and Technology, 08860 Castelldefels (Barcelona), Spain.}

\author{Maciej M. Ma\'ska}
\affiliation{Institute of Theoretical Physics, Wroc{\l}aw University of Science and Technology, 50-370 Wroc{\l}aw, Poland}

\author{Maciej Lewenstein}
\affiliation{ICFO - Institut de Ci\`encies Fot\`oniques, The Barcelona Institute of Science and Technology, 08860 Castelldefels (Barcelona), Spain.}
\affiliation{ICREA, Pg. Lluis Companys 23, ES-08010 Barcelona, Spain.}

\author{Tobias Grass}
\affiliation{DIPC - Donostia International Physics Center, Paseo Manuel de Lardiz{\'a}bal 4, 20018 San Sebasti{\'a}n, Spain}
\affiliation{Ikerbasque - Basque Foundation for Science, Maria Diaz de Haro 3, 48013 Bilbao, Spain}
\affiliation{ICFO - Institut de Ci\`encies Fot\`oniques, The Barcelona Institute of Science and Technology, 08860 Castelldefels (Barcelona), Spain.}

\author{Utso Bhattacharya}
\affiliation{ICFO - Institut de Ci\`encies Fot\`oniques, The Barcelona Institute of Science and Technology, 08860 Castelldefels (Barcelona), Spain.}

\begin{abstract}
Su-Schrieffer-Heeger (SSH) chains are paradigmatic examples of 1D topological insulators hosting zero-energy edge modes when the bulk of the system has a non-zero topological winding invariant.
Recently, high-harmonic spectroscopy has been suggested as a tool for detecting the topological phase. Specifically, it has been shown that when the SSH chain is coupled to an external laser field of a frequency much smaller than the band gap, the emitted light at harmonic frequencies strongly differs between the trivial and the topological phase. However, it remains unclear whether various non-trivial topological phases -- differing in the number of edge states -- can also be distinguished by the high harmonic generation (HHG). In this paper, we investigate this problem by studying an extended version of the SSH chain with extended-range hoppings, resulting in a topological model with different topological phases. We explicitly show that HHG spectra are a sensitive and suitable tool for distinguishing topological phases when there is more than one topological phase. We also propose a quantitative scheme based on tuning the filling of the system to precisely locate the number of edge modes in each topological phase of this chain.

\end{abstract}

\maketitle

\section{Introduction} 

In recent years, the field of condensed matter physics has seen an increasing interest in the study of topological phases of matter. These phases are characterized by non-local and non-perturbative properties that are protected by topological invariants, which are robust against perturbations and imperfections. The discovery of topological insulators~\cite{Hasan2010,asboth2016} and topological superconductors~\cite{Qi2011,viyuela2018} has led to the exploration of a wide range of topological phases in different materials, including cold atoms~\cite{atala2013} and photonic~\cite{rechtsman2013,stutzer2018,Ningyuan2015} systems. The study of these systems is not only of fundamental interest but also has potential applications in various fields, such as quantum computing~\cite{KITAEV20032,Plugge2017,Li2015,Li2014}, spintronics~\cite{cinchetti2014}, and magnetometry~\cite{nagaosa2013}.
The Su-Schrieffer-Heeger (SSH) model is one of the simplest models that exhibit non-trivial topological properties. It is a one-dimensional model originally introduced to describe the electronic properties of polyacetylene~\cite{Streitwolf1985,Block1996}, a linear polymer of carbon and hydrogen atoms. The model is described by a tight-binding Hamiltonian representing the hopping of electrons between adjacent sites, with two different parameters representing alternating single and double bonds. The SSH model exhibits two different phases, characterized by the number of edge states that appear in the band gap at zero energy. In the trivial phase, there are no zero-energy states, whereas, in the topological phase, there are two such states that appear at the open ends of the system. 
Various extensions of the SSH model have been explored, including longer-range tunneling terms that describe the hopping between second nearest neighbors~\cite{Li2014,Beatriz2019}. An appropriately extended SSH model may exhibit additional topological phases, such as a phase characterized by four edge states. 

High harmonic spectroscopy in condensed matter is a burgeoning field in strong-field attosecond science that has the potential to uncover the structural and dynamical properties of materials~\cite{Krausz2009,Calegari2016}. High harmonic generation (HHG) is a nonlinear optical process that occurs when an intense laser field interacts with a material, producing high-order harmonics of the incident frequency.
In recent years, the connection between strong-field attosecond science and topological condensed matter has started to be explored theoretically~\cite{Kelardeh2017,Bauer2018,Jurss2019,silva2019,Alexis2020,Drueke2019,Ikeda2018,Pattanayak2022,Baldelli2022} and experimentally~\cite{luu2018,reimann2018}.
In the context of the SSH model, theoretical studies have shown that high-harmonic spectroscopy can be used to detect topological properties~\cite{Bauer2018,Jurss2019}. In particular, the high-harmonic spectra of the SSH model exhibit characteristic features that allow one to distinguish between trivial and non-trivial topological phases and to identify the topological edge states.
Nevertheless, it remains unclear whether two different non-trivial topological phases, with various non-zero numbers of edge states, can also be identified using HHG.

To address this issue, in this paper, we consider the extended version of the SSH model, which includes second-neighbor electronic hopping as studied in Ref. ~\onlinecite{Beatriz2019}. This model exhibits topological phases with zero, two, and four-edge states. We
propose a method to distinguish between these phases using high-harmonic spectroscopy. Our method is based on the analysis of the non-linear polarization of the material, which reveals characteristic signatures of the topological phases. We show that our method can provide a clear distinction between materials with two- and four-edge states, and it can be used to identify and control the topological properties of other topological materials. Our method could have significant implications for the study of topological phases in condensed matter and could aid in the development of new technologies.

The paper has the following structure: In Section~\ref{sec:theory}, within Subsection~\ref{S2A}, we introduce the extended SSH model and discuss its band structure, topological properties, and phase diagram in comparison with the standard SSH model. In Subsection~\ref{S2B}, we provide specific details about the induced laser pulse and the coupling of light to the extended SSH chain. In Subsection~\ref{S2C} we shortly describe the numerical technique required to calculate the high harmonic spectrum for three different phases of the system. In Section~\ref{sec:results} we discuss the high harmonic spectra obtained for three different phases and investigate the role of filling in accurately determining the number of edge modes in each of the three phases of the system. Finally, we summarize our work in the concluding Section~\ref{sec:conclusions} with a brief outlook and experimental possibilities.

\section{Theory }\label{sec:theory}
	
\subsection{The extended Su-Schrieffer-Heeger model}\label{S2A}
The Su-Schrieffer-Heeger (SSH) model\cite{Su1979,meier2016} is a theoretical model used in solid-state physics to describe the electronic properties of one-dimensional crystalline systems. 
The model considers a chain of alternating atoms in sublattices A and B in a two-site unit cell~\cite{Jurss2019}. The electrons hop inside the unit cell (intra cell) and between nearest neighbor unit cells (inter cells) with different hopping amplitudes. The model is thus characterized by a parameter called the dimerization parameter, which represents the difference in the hopping strengths between the intra- and intercellular bonds. When the intercellular dimerization is stronger than the intracellular one, it exhibits a nontrivial topological phase with a bulk gap and supports the presence of topologically protected edge states at the boundary of the chain. This allows the SSH model to be a prototypical model of a 1D topological insulator. The topological nature of the SSH model is due to the presence of a chiral symmetry, which is a discrete symmetry that anti-commutes with the SSH Hamiltonian and shows that the model is invariant under the exchange of its two sublattices. It also ensures that for every positive energy of the system, there exists a negative energy with the same magnitude. Interestingly, the energies are also symmetric under swap of the dimerizations, and the dispersion relation is identical and gapped everywhere (insulator), except when the dimerization is zero, where it is gapless (metal). 
However, the system has distinct properties under swap of the dimerization as the eigenvectors differ significantly. In fact, when the Berry curvature of the eigenvectors in quasi-momentum space is integrated over the entire Brillouin zone, one finds different topological invariants called Chern numbers. This shows that the two insulating phases are topologically distinct. Different topological sectors imply the impossibility of crossing from an insulating phase to another without undergoing a topological phase transition, which involves the closing of the bulk gap (i.e. the metallic phase where the winding number is ill-defined) of the system. This is why when the system with a non-zero bulk topological invariant is put under an open boundary condition, there appears a zero-energy (edge) mode within the bulk gap of the system, sharply localized at the boundary separating a topologically non-trivial region (insulator) from a topologically trivial one. The non-interacting tight-binding model used in this study does not consider electron-electron interactions. As a result, it is possible to obtain precise analytical results for both the band structure and the winding number, making it a representative example of a 1D topological insulator~\cite{asboth2016}. So far, the SSH model has been experimentally realized in various systems: cold atoms~\cite{Lienhard2019}, photonic lattices~\cite{ Thatcher2022, Kanungo2022}, and mechanical systems~\cite{Rappoport2021}.

\begin{figure}[h]
\includegraphics[width=\columnwidth]{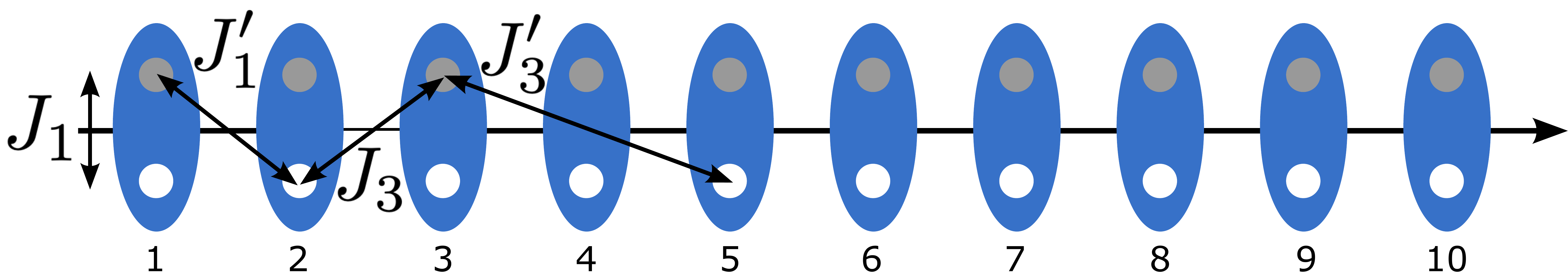}
\caption{\label{fig:cell_site_pos} Schematic representation of the one-dimensional ESSH model described by the Hamiltonian (Eq.~\eqref{H4}): Each blue ellipse represents a unit cell containing two sites: $A$-type sites drawn below and $B$-type sites drawn above. To keep chiral symmetry intact, we only include hopping processes between $A$-type and $B$-type sites, including intracell hopping, $J_1$, hoppings between neighboring cells $J_1'$ and $J_3$, as well as hopping between next-to-nearest cells $J_3'$.}
\label{fig:ESSH}
\end{figure}

The topological phase diagram of the standard SSH can be extended to include phases with higher values of topological invariants if not restricted to only nearest-neighbor electronic hopping. Allowing for a longer range of hopping, such as hopping between second neighbor sites, generates a model, which we denote as an extended Su-Schrieffer-Heeger (ESSH) model, extensively studied in Ref.~\onlinecite{Beatriz2019}. However, hoppings only between the different sublattices are included to preserve the chiral symmetry of the model, which in turn keeps the topology intact. More specifically, in this work, we study the Hamiltonian of the 1D ESSH model: 
\begin{equation}\label{eq:HSSHop}
\begin{split}
\mathcal{H}&= J_1 \sum^{N}_{n=1} (\hat{c}^{\dagger}_{n,A} \, \hat{c}_{n,B} +\mathrm{H.c.})  
+ J'_1 \sum^{N}_{n=1} (\hat{c}^{\dagger}_{n,B} \, \hat{c}_{n+1,A} +\mathrm{H.c.}) \\
&+ J_3 \sum^{N}_{n=1} (\hat{c}^{\dagger}_{n,A} \, \hat{c}_{n,B} +\mathrm{H.c.})
+ J'_3 \sum^{N}_{n=1} (\hat{c}^{\dagger}_{n,B} \, \hat{c}_{n+2,A} +\mathrm{H.c.}),
\end{split}
\end{equation}
where $N$ is the number of cells in a chain of $M=2N$ sites in the chain. In the second quantized notation, $\hat{c}^{\dagger}_{n,s}$ ($c_{n,s}$) is the electron creation (annihilation) operator in the unit cell $n$ with sublattices $s = A, B$. The first term represents intracellular electron hopping with strength $J_1$, the second represents the nearest neighbor intercellular hopping between $B$ in cell $n$ and $A$ in cell $n+1$, and the third is the nearest neighbor hopping between $A$ in cell $n$ and $B$ in cell $n+1$, while the fourth term captures the next nearest neighbor hopping between $B$ in cell $n$ and $A$ in cell $n+2$ (see Fig.~\ref{fig:ESSH}). 
In the first quantized notation, the Hamiltonian in real space can be written as
\begin{equation}\label{H4}
\begin{split}
\mathcal{H}&= J_1 \left(\sum^N_{n=1}|n,A \rangle \langle n. B |  + \mathrm{H.c.} \right) \\
&+ J'_1 \left(\sum^N_{n=1}|n, B \rangle \langle n+1,A |  + \mathrm{H.c.} \right)\\ 
&+ J_3 \left(\sum^N_{n=1}|n,A \rangle \langle n+1,B | + \mathrm{H.c.} \right) \\
&+ J'_3 \left(\sum^N_{n=1}|n,B \rangle \langle n+2,A |  + \mathrm{H.c.} \right). 
\end{split}
\end{equation}

As in the standard SSH model, the extended SSH model Hamiltonian also preserves the three discrete symmetries: chiral, particle-hole, and time-reversal, is classified under the BDI topological class with its topological invariant (the winding number) belonging to the set of integers $\mathbb{Z}$. On an open boundary, the ESSH can host different numbers of edge modes: zero, two, or four edge modes, depending on the absolute value of its bulk winding number is zero, one, or two. In contrast, the standard SSH model possesses only two possibilities -- zero or two edge modes on the ends of the open chain.

\begin{figure*}[ht!]
\includegraphics[width=1.9\columnwidth]{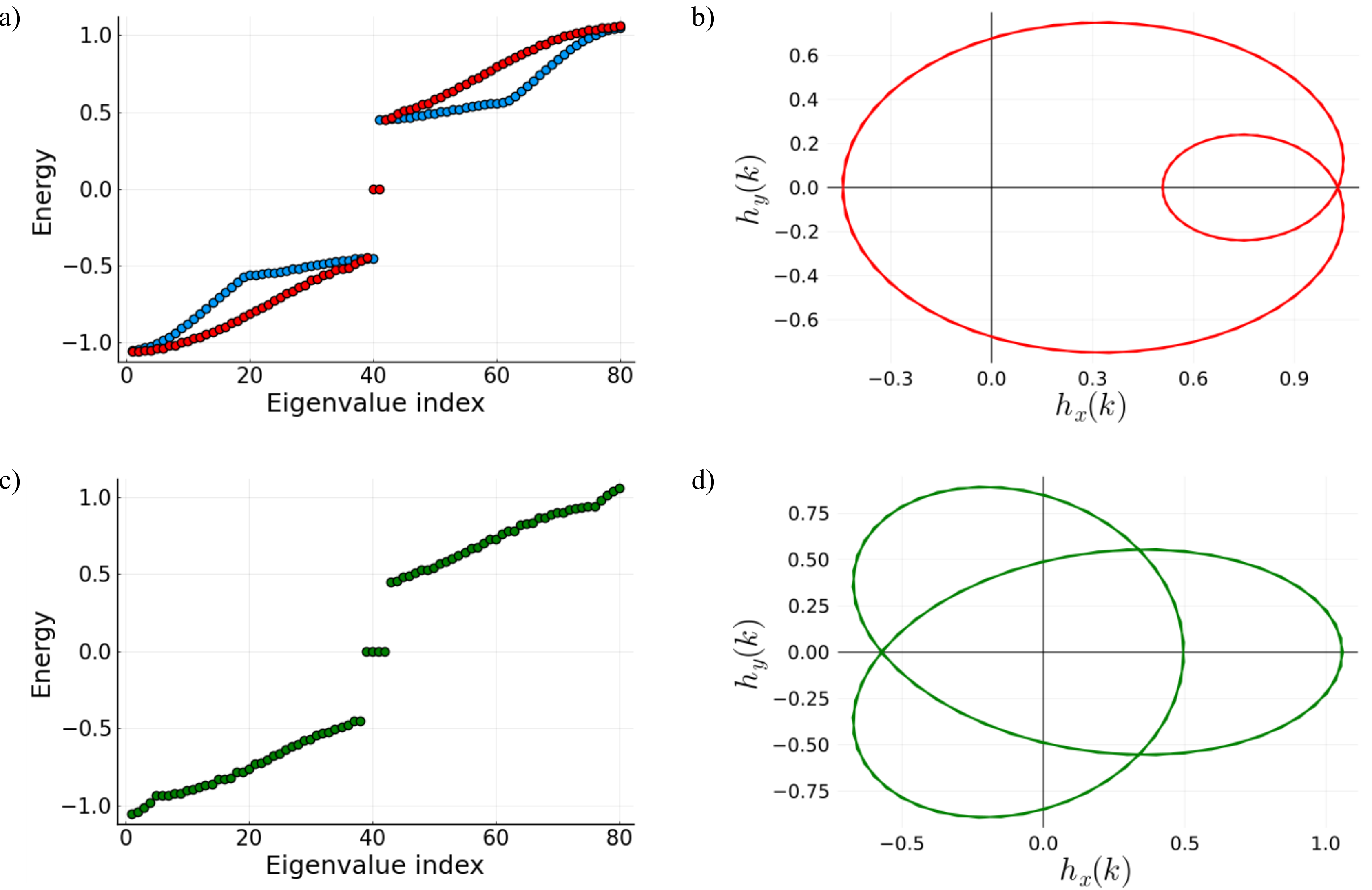}
\caption{Energy eigenvalues with $80$ sites (in a.u.), showing zero (a, blue), two (a, red), and four (c, green) zero-energy states with different fixed parameter values of the Hamiltonian (Eq.~\eqref{H4}). The figures on the right show the parametric plot of $h_x (k)$ and $h_y (k)$ as defined in Eq.~\eqref{eq:Hmomentum}, with two (b) and four (d) zero energy states.}
\label{fig:Espectrum}
\end{figure*}

\begin{figure}[ht!]
\centering
\includegraphics[width=\columnwidth]{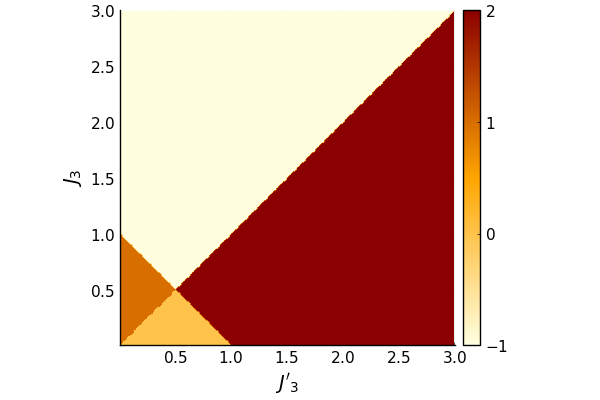}
\caption{Topological phase diagram of the Hamiltonian (Eq. \eqref{eq:Hmomentum}) showing various values of the winding number for fixed $J_1 =J'_1$ varying $J_3$ and $J'_3$ (in units of $J_1$), calculated as in Eq. \eqref{eq:Wnumber}. }
\label{fig:Wn}
\end{figure}

It is straightforward to calculate the winding number by writing down the Hamiltonian in Eq.~\eqref{H4} in momentum representation, which can be obtained by replacing,
\begin{equation}
\begin{split}
|n,A\rangle&= \sum_{k} e^{- i k x_{nA}}|k,A\rangle, \\
|n,B\rangle&= \sum_{k} e^{- i k' x_{nB}}|k', B\rangle,
\end{split}
\end{equation}
where $|k, s\rangle$ is the quasi-momentum ket with momentum $k$ and sublattice index $s= A, B$ and using periodic boundary conditions. In the rest of the work, we denote the lattice constant by $a$. With this, the Hamiltonian in Eq.~\eqref{H4} reduces to
\begin{equation}
\label{eq:Hmomentum}
\mathcal{H}= \sum_{k}\left(
 | k,A \rangle , | k,B \rangle\right)  
 \left[ h_x(k) \sigma_x + h_y(k) \sigma_y \right]
 \begin{pmatrix}
 \langle k,A | \\
\langle k,B | 
\end{pmatrix},
\end{equation}
where 
\begin{eqnarray}
    h_x(k) &=& J_1 + J'_1\cos{ka}+ J_3\cos{ka}+ J'_3\cos{2ka}, \nonumber \\
    h_y(k) &=& J'_1\sin{ka}- J_3\sin{ka}+ J'_3\sin{2ka}.
\end{eqnarray}
Due to the presence of discrete translational invariance in the system, it is reduced to two-level systems in the sublattice basis for each quasi-momentum mode $k$ and can therefore be easily written down in terms of Pauli matrices $\sigma_x$ and $\sigma_y$. Furthermore, this decomposition into the Pauli basis allows us to compute the winding number using $h_x$ and $h_y$,
\begin{align}\label{eq:Wnumber}
    \mathcal{W}= \frac{1}{2\pi} \int\limits_{BZ}\frac{h_x \partial_{k} h_y - h_y \partial_{k} h_x }{h_x^2 +h_y^2} \mathrm{d}k.
\end{align}
The winding number for this model can have values $\mathcal{W}=-1,0,1,2$, which through the bulk-boundary correspondence directly yields the number of edge modes the system possesses in an open boundary condition as two times its absolute value $2\vert\mathcal{W}\vert$. From the energy spectrum plotted in Fig.~\ref{fig:Espectrum}, one can see the band gap structure, the different number of zero energy modes or edge modes, and the winding number for the two most interesting cases when the system (a) has winding number one and thus one pair of edge modes, and (b) has two pairs of edge modes with winding number two. Fixing the value of the parameters $J_1 = J'_1 = 1$, one can vary the other two parameters $J_3$ and $J'_3 $ and calculate the winding number to obtain a phase diagram showing different possible phases in the ESSH system, as illustrated in Fig.~\ref{fig:Wn}.

\subsection{Incident laser field} \ \label{S2B}
In this section, we study the coupling of the 1D ESSH model to a linearly polarized electric field from a laser. The laser wavelength is assumed to be much larger than the length of the system, and as such, the coupling to the laser field is well captured within the dipole approximation. The laser vector potential and electric field are:
\begin{equation}\label{E_A} \vec{A}(t) =
A(t)\hat{x},~~~~~\vec{E}(t) = -\partial_t\vec{A}(t).
\end{equation}
where $\hat{x}$ is the direction along the length of the chain parallel to the laser polarization. 
The way in which the light couples to the matter depends on the geometry of the system. For instance, in the velocity gauge, the light-matter coupling provides the hopping elements with Peierls' phases ${\vec A}\cdot ({\vec r}_{n,s}-{\vec r}_{n',s'})$. For concreteness, we assume that ${\vec A}\cdot ({\vec r}_{n,s}-{\vec r}_{n',s'}) \propto n-n'$. Therefore, the intracell hopping remains unaffected by the light, i.e. ${\vec A}\cdot ({\vec r}_{n,A}-{\vec r}_{n,B})=0$, whereas hopping between neighboring cells acquires a phase $A(t)$, and hopping between next-to-nearest cells a phase $2A(t)$. Accordingly, we have
\begin{equation}
\begin{split}
J_1 (t) =& J_{1}, \qquad \qquad \; \; J'_1 (t) = J'_{1} e^{i a A(t)},\\
J_3 (t) =& J_{3} e^{i a A(t)}, \qquad J'_3 (t) = J'_{3} e^{2 i  a A(t)}.\\
\end{split}
\end{equation}

The eigenstates of the $N\times N$-dimensional ESSH Hamiltonian~\eqref{eq:HSSHop} are obtained by exact diagonalization in a real space single particle basis. The $N/2$ lowest energy states (occupied by $N$ electrons, assuming spin
degeneracy) are time-evolved within the whole laser pulse duration consisting of $5$ cycles ($n_c=5$). Assuming atomic units ($\hbar=|e|=m_e=4\pi\epsilon_0=1$), the incident laser field has the shape
\begin{equation}\label{eq:Vpot}
A(t) = A_0 \sin^2\left(\frac{\omega t}{2 n_c}\right) \sin\left(\omega t\right),\qquad 0<t<\frac{2\pi n_c}{\omega}.
\end{equation}
The frequency is set to $\omega=0.03$ (corresponding to $\lambda \simeq 1.5{\rm \, \mu m}$), and the vector potential amplitude is $A_0 = 0.5$ (corresponding to a laser intensity $\simeq 20 \times 10^{10} {\rm  \, Wcm^{-2}}$ )
throughout the paper. The results which we discuss in this paper do not depend on the details of the laser pulse as long as the incident laser frequency is small compared to the band gap in the insulating phases, and the peak strength of the laser is large enough to generate high harmonics.

The evolution of the wavefunction with the time-dependent Hamiltonian was calculated using the Crank-Nicolson approximation, 
\begin{align}  \label{eq:CN}
\ket{\Psi(t+\delta t)}&=\exp[-i\,\mathcal{H}(t)\delta t] \ket{\Psi(t)}\nonumber \\ 
&\approx \frac{1-i\,\mathcal{H}\left(t+\delta t/2\right)\delta t/2}{1+i\,\mathcal{H}\left(t+\delta t/2\right)\delta t/2} \ket{\Psi(t)},
\end{align}
solved in individual infinitesimal $\delta t$ time steps {, with the initial condition $\ket{\Psi(0)}$ being the ground state of the system.}

\subsection{High-harmonic generation}
\label{S2C}
Our primary goal is to estimate the high harmonic spectrum of the ESSH system. Within the semi-classical approach, for uncorrelated emitters, the spectrum
of the radiated light is proportional to the absolute square of the Fourier transform of the dipole acceleration~\cite{Bandrauk2009,Baggesen2011,Sundaram1990},
\begin{equation} \label{eq:power}
P(\omega) = \left| \mathrm{FFT}\left[
W_B\ddot{X}(t)\right]\right|^2,
\end{equation}
where we used $W_B$ as the Blackman window function~\cite{Podder2014} and $\ddot{X}(t)$ is the acceleration or the double time derivative of the time-dependent position operator.
To calculate the time-dependent expectation value of the position operator, we time-evolve all occupied eigenstates $b$,  $b \in (1, N/2)$ 
The time-evolved single-particle wavefunction $\Psi_b(t)$, is then used to compute
\begin{equation}\label{eq:Xtfilled}
X(t) = \sum_{b = 1}^{N/2}\sum_{j=1}^{N}\sum_{s=A,B} \Psi_b^{j,s*}(t) x_{j,s} \Psi_b^{j,s}(t),
\end{equation}	
where $\Psi_b^{j,s}(t)$ is the amplitude of the time evolved wavefunction on site $s$ of cell $j$, and the position $x_{j,s}$ is given by
\begin{equation}
x_{j,A} = x_{j,B} = (j-1) - \frac{(a-M)}{4},
\end{equation}
where $j$ is the position of each cell with two sites $A$ and $B$, as shown in Fig.~\ref{fig:cell_site_pos}. Essentially, the time-evolved average positions of all electrons in different (initial) filled eigenstates are summed to obtain the total position of the electron cloud.

In the next Section, we analyze and compare the time-dependent position and the harmonic response of the system at three different parameter points, corresponding to three different phases as illustrated in Fig.~\ref{fig:Espectrum}: (a) Phase $P_0$, represented by the parameter point $J_1= 0.651$, $J'_1= 0.207$, $J_3 = 0.038$, and $J'_3 = 0.156$. From the energy spectrum presented in Fig.~\ref{fig:Espectrum}a (blue), we observe the energy band gap of $0.9$ (in atomic units) and the winding number of $\mathcal{W}=0$ corresponding to the trivial insulator phase. (b) Phase $P_1$, represented by the point $J_1= 0.51$, $J'_1 = 0.42$, $J_3 = 0.056$, and $J'_3= -0.479$. It is illustrated in Fig.~\ref{fig:Espectrum}a (red), exhibiting the presence of two zero-energy states that indicate the edge states, in agreement with the winding number $\mathcal{W}=1$. (c) Phase $P_2$, represented by $J_1 = 0.059$, $J'_1 = 0.021$, $J_3 = 0.26$, and $J'_3 = 0.7209$. For this phase, we observe in Fig~\ref{fig:Espectrum}b the presence of four zero-energy states, representing the edge modes with $\mathcal{W}=2$. The parameter choices have been made such that the system has the same band gap and bandwidth in all three phases, allowing for a clear comparison of the results amongst all three phases. We also look at the phase transition point $M$, which is a metal with the parameters $J_1=0.51$, $J'_1=0.42$, $J_3=1$, and $J'_3=0.91$. The overall goal is to observe whether, through the time-dependent position operator and the harmonic spectra, one can identify certain signatures, which will enable a clear distinction between $P_0, P_1, P_2$, and $M$. 

\section{Results and Discussions}\label{sec:results}
\begin{figure}[ht!]
\centering
\includegraphics[width=\columnwidth]{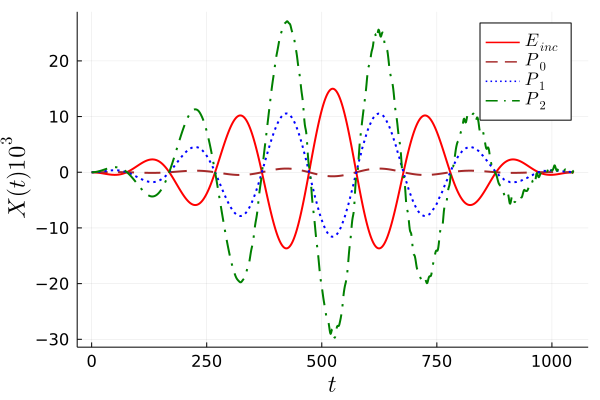}
\caption{Incident electric field (red, solid line) and the expected value of the position operator (Eq.\eqref{eq:Xtfilled}) as a function of time for different topological phases $P_0$, $P_1$, and $P_2$.}
\label{fig:Xt}
\end{figure}

We first look at the expectation value of the total position operator of the ESSH model as a function of time, as illustrated in Fig.~\ref{fig:Xt} for the three different phases $P_0, P_1$, and $P_2$. The position operator has a similar periodicity in all the phases as that of the incident laser beam (Eq.~\eqref{eq:Vpot}), but there is a clear difference in the maximum amplitude for the three phases. Apparently, the change in the average electronic position is the lowest in the topologically trivial insulating phase $P_0$ and can be attributed to the overall localized nature of the electronic cloud for a half-filled insulator. The presence of edge modes makes the system slightly more metallic, and hence the displacement is more in phases ($P_1, P_2$) with more edge modes in the system. This becomes more apparent as one studies the harmonic spectra of the system.

\begin{figure}[ht!]
\centering
\includegraphics[width=\columnwidth]{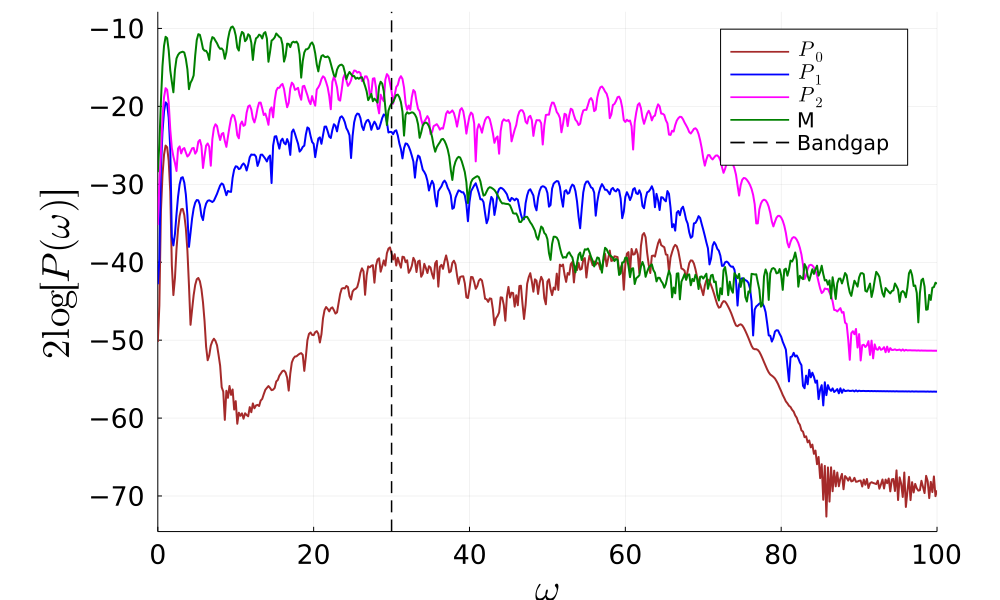}
\caption{The high-harmonic spectra emitted for the phases: metallic M, topologically trivial $P_0$, and two different topologically non-trivial $P_1$ (with two edge modes) and $P_2$ (with four edge modes). The vertical line corresponds to the value of the bandgap (in the units of the incident laser frequency).}
\label{fig:HHG}
\end{figure}

In Fig.~\ref{fig:HHG}, we plot the logarithm of the absolute power spectra of the harmonic spectra versus the harmonic order (integer multiple of the incident driving frequency). 
The harmonic spectra for $P_0, P_1$ and $P_2$ show a plateau at high energies beyond the bandgap of the system. The plateau arises mainly from interference between electronic trajectories that undergo interband transitions.  A cut-off is also observed at similar harmonic order for all three curves, as it is primarily determined by the bandwidth (energy difference between highest and lowest eigenenergy) of the system, which limits the maximum energy that the electrons can attain during evolution. However, the harmonic response below the band gap is different for the $P_0$ and $P_1, P_2$ phases. This region, which in usual semiconductors is mainly dominated by intraband contributions, has a dip for $P_0$. However, in phases $P_1$ and $P_2$, we do not observe a dip below the band gap, because mid-band gap states are available for electronic transition due to edge modes. These appear as clear signals in the harmonic spectra as now transitions between the filled bulk bands to the mid-gap edge states are possible. In contrast, the $P_0$ phase is a trivial insulator with a bulk gap and no mid-gap states, and thus the bulk states can only contribute to the harmonic spectra beyond the band gap. Consequently, there occurs a dip in the signal below the band gap for such a phase. This feature that allows one to distinguish between the high harmonic spectra of topologically trivial and non-trivial phases was also observed in previous studies of the HHG in the SSH model\cite{Bauer2018,Jurss2019}. Moreover, it has been studied in topological superconductors, and the topological nature of the edge modes has been confirmed by showing that it is robust under local perturbations via the HHG\cite{Baldelli2022}.

However, interestingly, it is not easy to distinguish between the two topological phases based on the harmonic spectra itself, as the overall amplitude difference in the time-dependent position being polynomial, does not appear as a big difference in the harmonic spectra, which is plotted on a logarithmic scale and depends on the specific values of the hopping parameters chosen for these two phases.
Despite this, by analyzing the contributions of both bulk and edge states to the high-harmonics and using the harmonic spectra of the trivial insulating phase $P_0$ and the metallic phase transition point $M$ as two extreme reference limits to test for metallicity, we can elaborate below how precise control over the electronic filling in the ESSH chain allows us to clearly distinguish between all the different topological phases based solely on the HHG spectra.

We assume an ESSH chain away from half-filling where the number of electrons in the chain is $\nu$ less than $N/2$. Then the expectation value of the position operator is given by
\begin{equation}
X(t,\nu) = \sum_{b = 1}^{N/2-\nu}\sum_{j=1}^{N}\sum_{s=A,B} \Psi_b^{j,s*}(t) x_{j,s} \Psi_b^{j,s}(t).
\end{equation}	
As a consequence, varying the ESSH chain filling, i.e., changing the value of $\nu=\{0,1,2,3,4\}$ affects the HHG spectra (Eq.~\eqref{eq:power}) of phases $P_0$, $P_1$, and $P_2$ as shown in Fig.~\ref{fig:Espectrum}.

We first focus on phase $P_0$ (see Fig.~\ref{fig:Espectrum}a) and compare the HHG spectra for various fillings against the one at exactly half-filling (yellow). It can be clearly observed that for filling values up to $N/2-1$, the HHG spectra are almost identical regardless of the filling. In fact, away from half-filling, the spectra have no dips at half the band gap. This is because even slightly away from half-filling, the system is no longer an insulator, and there are a small number of states available (depending on the filling) for transition within the bulk states below the band gap, producing significant HHG spectra from intraband dynamics within this partially filled valence band. These HHG spectra resemble those of a metal, as can be seen by comparison with the green curve in Fig.~\ref{fig:HHG}. The order-of-magnitude difference (in logarithmic scale) between the spectra at half-filling and the one away shows how sensitive the probe the HHG spectra is to the filling of the system that produces metallicity and that does not. 

This sensitivity acts as a means of quantitatively distinguishing the presence of the number of edge modes in phases $P_1$ (see Fig.~\ref{fig:Espectrum}b) and $P_2$ (see Fig.~\ref{fig:Espectrum}c). By varying the filling of the system with parameters from the $P_1$ ($P_2$) phase, we see that the system shows a metallic HHG spectra till $N/2-2$ ($N/2-3$) states are filled, then suddenly it shows a dip as the bulk of the system becomes insulating when $N/2-1$ ($N/2-2$) states are filled. Thus, by continuously monitoring the filling, it is possible to count how many states ahead of half-filling the system show a transition from metallic behavior to insulating one. This difference in the number of states gives the number of pairs of edge modes in the system.

The dip in the harmonic power spectra at the filling, where the transition from bulk metallic to bulk insulating behavior occurs, can be quantitatively determined by summing the inverse of the squared value of power spectra below half the bandgap for every value of $\nu$ as,
\begin{equation} \label{eq:filling}
S_p(\nu)=\sum_{w=0}^{\Delta E/2} \frac{1}{P(\omega,\nu)^2},
\end{equation}
where $\Delta E$ is the bulk band gap energy of the system (edge modes excluded). The sum is taken over half the band gap; as expectedly, the sharp change only affects the harmonic modes below the band gap in the harmonic spectra.

\begin{figure}[ht!]
\centering
\includegraphics[width=\columnwidth]{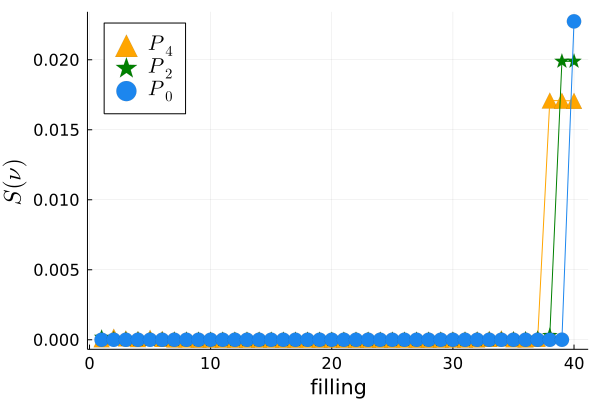}
\caption{ $S_p(\nu)$ versus filling for different phases $P_0$, $P_2$ and $P_4$. It shows a peak when all the states in the bulk are filled and the edge states start to be filled. One state in the peak represents zero edge modes or Phase $P_0$ ($S_{P_0}(\nu) \sim 10^9 S_{P_4}(\nu)$), two states in the peak represent two edge modes or Phase $P_2$ ($S_{P_2}(\nu)\sim 10^7 S_{P_4}(\nu)$ ) and three states in the peak represent four edge modes or Phase $P_4$.}
\label{fig:Snu}
\end{figure}

The transition shows up in this quantity $S_p(\nu)$ as a sharp jump with at least ten orders of magnitude difference. In phase $P_0$, the transition occurs between the completely filled, $\nu=0$, to $\nu=1$, where $S_p(0)\sim 10 S_p(1)$ says that the number of pairs of edge modes in the system is zero. In phase $P_1$, the transition takes place between $\nu=1$ to $\nu=2$, where $S_p(0)\sim S_p(1)\sim 10 S_p(2)$ and then the system possesses just one pair of edge modes. In phase $P_2$, the transition is from $\nu=2$ to $\nu=3$, which changes $S_p(\nu)$ as $S_p(0)\sim S_p(1)\sim S_p(2)\sim 10 S_p(3)$ correctly indicating that there are two pairs of edge modes in the system. We show an illustrative plot of this behavior in Fig.~\ref{fig:Snu}. 

\begin{figure*}[ht!]
\includegraphics[width=2\columnwidth]{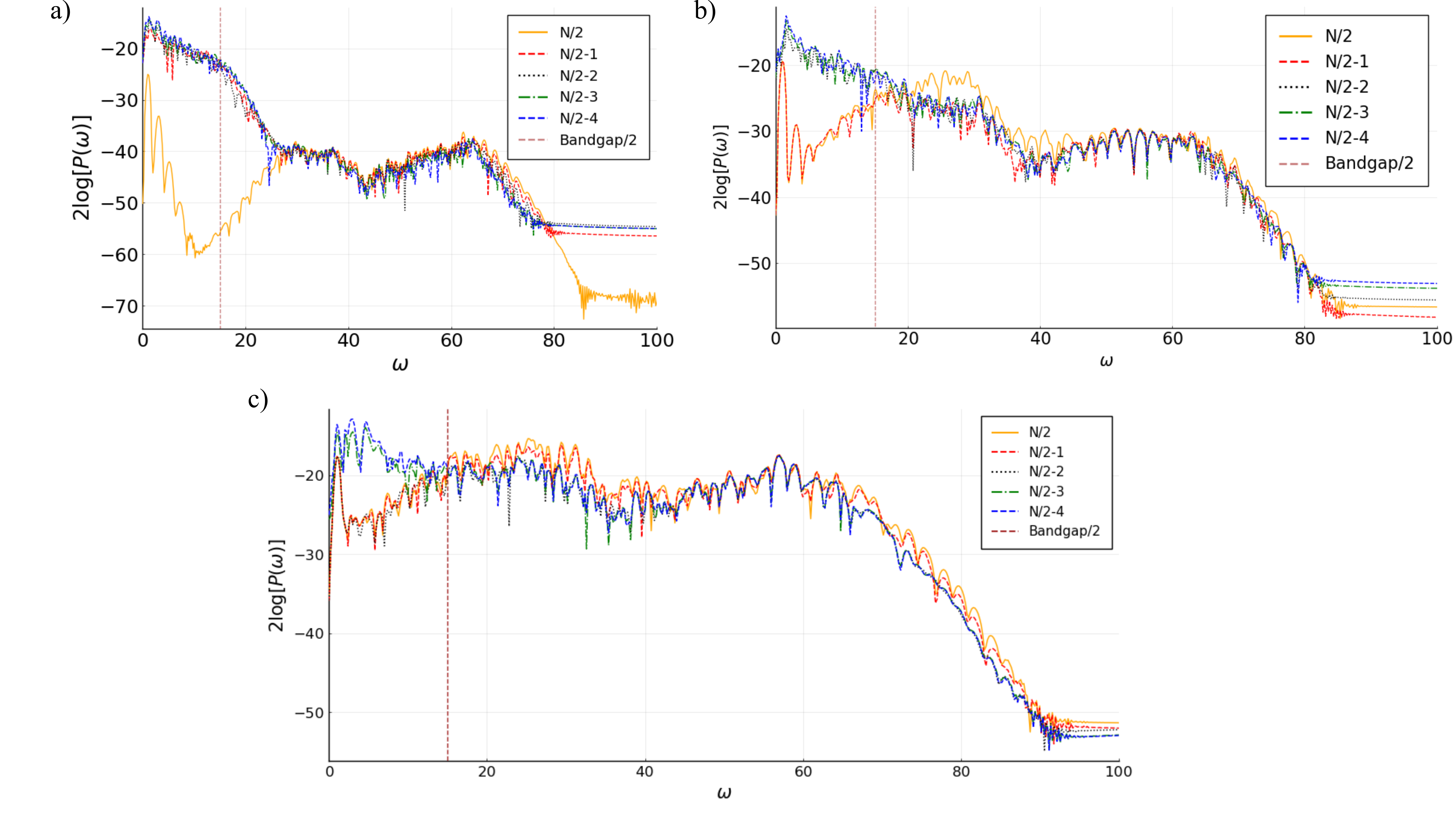}
\caption{High-harmonic spectra for different phases. a) Phase $P_0$ (zero edge modes). b) Phase $P_1$ (two edge modes).  c) Phase $P_2$ (four edge modes). Different colors correspond to various fillings of the system. The vertical line indicates half of the value of the bandgap (in units of the incident laser frequency).}
\label{fig:HHspectrum}
\end{figure*}

\section{Conclusions}\label{sec:conclusions}
 In summary, in this work, we allow for second nearest-neighbor hopping in addition to the nearest-neighbor hopping already present in the standard SSH model, with hopping terms within a sublattice being forbidden. This creates the ESSH model with the chiral symmetry in the SSH model being preserved. However, this expands the topological phase diagram of the SSH model to now include new topological phases with higher winding numbers. Such a system under an open boundary condition has three insulating phases with zero, two, and four edge modes at each end of the chain. 
 
 We shine a five-cycle ultrafast laser pulse with strong intensities and below the bulk band gap frequencies parallel to the length of the ESSH chain and calculate the emitted harmonic spectra in response to this illumination. The harmonic spectra in the linear scale shows that the overall below-the-band gap response of the system is different for its three phases, the amplitude being higher when the system has more edge modes. However, although the distinction is clearly visible between the HHG spectra from the trivially insulating versus the topological ones, the distinction between the two topological phases is hard to perceive in the logarithmic scale. Therefore, we did a careful analysis of the HHG spectra as a function of filling to show that the HHG spectra are very sensitive to the change from bulk insulating behavior to bulk metallic behavior, as filling is continuously varied. Therefore, tracking where the bulk insulating behavior sets in, as a function of filling, we manage to count off the number of pairs of edge states in the system. We have also proposed a quantity that can sharply detect this transition.

Our work concentrates on studying an idealized model which features different topological phases. Our goal was to investigate whether and how HHG is suitable for distinguishing between these phases. Of course, the next step will be to consider the analog phases in real materials.  This brings in additional challenges: In the present work, we have not included the effect of scattering between electrons or electron-phonon and other defects. A phenomenological way to consider such effects is by including a dephasing time in the analysis \cite{Alcala2022}. This can help produce a cleaner spectrum by removing longer trajectories contributing to the HHG spectra. We leave this as an outlook. In addition, the role of the many-body electron-electron interaction has been assumed to be negligible, and this has not been considered in this work.

{Finally, as an outlook, we would like to point out that proposed HHG spectroscopy can serve as a tool for topological phase detection in 2D topological superconductors  classified by an
integer-valued Chern number.
In topological superconductors, the non-trivial topology is related to
the quantization of electronic Hall thermal conductance \cite{Hasan_2015, Sato_2017,Sharma_2022}. However, thermal conductance
measurements have not reached the
required sophistication to observe the quantization, and recently only machine learning approaches to topological invariants detection  \cite{baireuther2021identifying} were proposed. As such, HHG spectroscopy is a promising tool for Chern number identification in topological superconductors.}

\begin{acknowledgments}

ICFO group acknowledges support from: ERC AdG NOQIA; Ministerio de Ciencia y Innovation Agencia Estatal de Investigaciones (PGC2018-097027-BI00/10.13039/501100011033,CEX2019-000910-S/10.130 39/501100011033, Plan National FIDEUA PID2019-106901GB-I00, FPI, QUANTERA MAQS PCI2019-111828-2, QUANTERA DYNAMITE PCI2022-132919,  Proyectos de I+D+I “Retos Colaboración” QUSPIN RTC2019-007196-7); MICIIN with funding from European Union NextGenerationEU(PRTR-C17.I1) and by Generalitat de Catalunya;  Fundació Cellex; Fundació Mir-Puig; Generalitat de Catalunya (European Social Fund FEDER and CERCA program, AGAUR Grant No. 2021 SGR 01452, QuantumCAT \ U16-011424, co-funded by ERDF Operational Program of Catalonia 2014-2020); Barcelona Supercomputing Center MareNostrum (FI-2022-1-0042); EU (PASQuanS2.1, 101113690); EU Horizon 2020 FET-OPEN OPTOlogic (Grant No 899794); EU Horizon Europe Program (Grant Agreement 101080086 — NeQST), National Science Centre, Poland (Symfonia Grant No. 2016/20/W/ST4/00314); ICFO Internal “QuantumGaudi” project; European Union’s Horizon 2020 research and innovation program under the Marie-Skłodowska-Curie grant agreement No 101029393 (STREDCH) and No 847648(“La Caixa” Junior Leaders fellowships ID10001 0434:LCF/BQ/PI19/11690013, LCF/BQ/PI20/11760031,LCF/BQ/PR20/11770012,LCF/BQ /PR21/11840013). MLB acknowledges the financial support from MCIN/AEI/10.13039/5011000 11033. M.M.M. and M.D. acknowledge support from the National Science Centre (Poland) under Grant No.DEC-2018/29/B/ST3/01892. M.P. acknowledges the support
of the Polish National Agency for Academic Exchange, the Bekker programme no: PPN/BEK/2020/1/00317. Views and opinions expressed are, however, those of the author(s) only and do not necessarily reflect those of the European Union, European Commission, European Climate, Infrastructure and Environment Executive Agency (CINEA), nor any other granting authority.  Neither the European Union nor any granting authority can be held responsible for them. 

\end{acknowledgments}
\appendix

\bibliography{HHGESSH.bib}	
\end{document}